\documentstyle[bibnorm,psfig]{lamuphys}

\begin{document}

\title{Mechanical Oscillations at the Cellular Scale}
%\title{Mechanical oscillators in living cells}
\author{Frank J\"ulicher}
\institute{Institut Curie, Physicochimie Curie, 
Section de Recherche, 26 rue d'Ulm,
75248 Paris Cedex 05, France}
\maketitle

\begin{abstract}
    Active phenomena which involve force generation and motion play a
    key role in a number of phenomena in living cells such as cell
    motility, muscle contraction and the active transport of material
    and organelles.  Here we discuss mechanical oscillations generated
    by active systems in cells.  Examples are oscillatory regimes in
    muscles, the periodic beating of axonemal cilia and flagella and
    spontaneous oscillations of auditory hair cells which play a
    role in active amplification of weak sounds in hearing.  As a
    prototype system for oscillation generation by proteins, we
    discuss a general mechanism by which many coupled active elements
    such as motor molecules can generate oscillations.
\end{abstract}

\begin{abstract}
    Les ph\'enom\`enes actifs impliquant la g\'en\'eration de forces
    et de mouvement, jouent un r\^ole primordial dans de nombreux
    ph\'enom\`enes de la vie cellulaire tels que la motilit\'e, la
    contraction musculaire et le transport actif d'organelles ou de
    compos\'es. Nous discutons ici les oscillations m\'ecaniques 
    g\'en\'er\'ees par des syst\`emes actifs dans les cellules. Des
    exemples sont les r\'egimes oscillatoires des muscles, le battement
    p\'eriodique des cils et des flagelles axonemaux et les oscillations
    spontan\'ees des cellules auditives. Ces oscillations spontan\'ees
    ci pourraient jouer un r\^ole important dans l'audition par
    l'amplification des faibles sons. Nous discutons un syst\`eme 
    prototype de g\'en\'eration d'oscillations par des prot\'eines,
    en d\'ecrivant un m\'ecanisme g\'en\'eral impliquant l'intervention
    de nombreux \'el\'ements actifs, comme des moteurs 
    mol\'eculaires.\\ \\
    Mots cl\'es: biophysique/motilit\'e/oscillations/audition/axoneme
\end{abstract}

\section{Introduction}

The generation of forces and the ability to move represent some of the
most striking abilities of living cells.  Prominent examples are cell
motility, the contraction of muscles but also active transport of
materials and of organelles, for example during cell division and
mitosis.  Such movements are generated on the molecular level by
protein molecules that convert chemical energy to mechanical work. 
Prominent examples are linear motor proteins of eucariotic cells.  These
motors are specialized to work by interacting with filaments of the
cytoskeleton \cite{krei93}.  They consume Adenosintriphosphate (ATP)
as a fuel and convert its chemical energy to mechanical work.  Myosin
motors for example generate motion along actin filaments. They are the
active components in muscles, while kinesin and dynein motors move along
microtubules \cite{albe94}.

In certain situations, cells can generate oscillatory motion and
operate as mechanical oscillators.  Oscillatory behaviors of muscles
have been known for a long time in the case of some flight muscles
which move the wings of insects \cite{prin77}.  Many insects such as
wasps and bees have muscles called "asynchronous" which generate
oscillatory contractions by a mechanism located in the muscle itself
which drives the beating of the wings.  This type of muscle differs
from "synchronous" flight muscles of some other insects for which
contractions are triggered by a periodic nerve signal and are not
generated in the muscle.  More recently, oscillatory behaviors have
been observed under certain conditions in fibrils of ordinary skeletal
muscles which under normal conditions do not oscillate 
\cite{yasu96,fuji98}. 

An important example for mechanical oscillations on the cellular scale
is the periodic motion of cilia and flagella \cite{bray92,mura92}.  These
hair-like appendages of many cells are used for swimming and
self-propulsion by sperm and some small organisms or to stir fluids
surrounding a cell or cell layer.  The common structural feature of
these cilia and flagella is the axoneme, a well conserved arrangement
of microtubule doublets organized in a cylindrical fashion.  Dynein
molecular motors are attached to these microtubules and can exert
forces on neighboring microtubules.  This structure is very well
conserved and appears in a large number of different cells of
different organisms \cite{lind97}.  The activity of the dynein
molecular motors coupled to the microtubules leads to periodic bending
deformations and waves along the cilium.  In the case of sperm, the
flagellum propagates bending waves from the head towards the tail
\cite{gibb75}.  These waves are very regular and essentially planar
\cite{brok91,brok96} and break the necessary symmetries to allow the
sperm to swim in a viscous solution \cite{tayl51,purc77}.

In the following sections, we describe some properties of mechanical
oscillators and their implications for biological situations.  Close
to an oscillating instability called Hopf-bifurcation \cite{stro94},
the behavior of a mechanical oscillator and its response to externally
imposed oscillating forces are generic \cite{cama00,egui00}.  As a
prototype system for the generation of oscillations in cells, we
discuss a general mechanism by which a large number of active elements
such as molecular motors which are coupled to an elastic element and
which undergo a chemical cycle of fuel consumption can generate
mechanical oscillations \cite{juli97,juli97b}.  These oscillations are
a result of collective behavior of many motors \cite{huxl57,juli95}
which can give rise to different types of dynamic transitions
\cite{juli97,rive98,reim99,reim00}.  The same physical mechanism can
also generate oscillatory modes and bending waves along elastic
filaments which interact with many motors as in the case of cilia
\cite{brok75,cama99,cama00b}.  Finally, oscillating instabilities can
play an important role in sensory systems.  A lot of evidence points
to an active system involved in the detection of sound by the ear
\cite{gold48,dall92,huds97,zhen00}.  The inner ear shows a frequency
selective nonlinear response to sound stimuli and actively amplifies
weak sounds \cite{rugg92}.  These properties allow the ear to cover a
large dynamic range of $120$dB and to detect weak sounds which per
cycle of oscillation impart an energy that is less than kT
\cite{bial87}.  These properties of the ear can be understood if we
assume that the inner ear contains dynamical systems which operate at
Hopf bifurcation with critical frequencies that cover the audible
range.  A simple self-tuning mechanism involving a feedback control
can explain how these systems tune reliably to their critical point. 
The concept of a self-tuned Hopf bifurcation can explain many
apparently distinct phenomena observed in hearing such as otoacoustic
emissions, nonlinearities, adaptation and fatigue as well as the
response to complex sounds in a unified framework \cite{cama00,juli01}.

\section{Mechanical oscillators}

A large variety of nonlinear dynamic systems is capable to generate
periodic, oscillating motion. In order to
illustrate some of the general properties of mechanical oscillators, we consider
a dynamical system related to the Van der Pol oscillator which is 
standard model for nonlinear oscillators \cite{vand26,nayf79}.  
Consider the dynamic equations
\begin{equation}
    \gamma \ddot x + r \dot x - \Lambda \dot x^3 + k x = f_{\rm 
    ext}(t)\quad,\label{eq:vdp}
\end{equation}
which represent a damped oscillator with additional nonlinear
friction.  Here $x$ is a displacement variable and $f_{\rm ext}$
an external force.  We assume that $\gamma$, $k$ and $\Lambda$ are
positive parameters and that $r$ can become negative. Note that 
while the term characterized by $\gamma$ could be an inertial term,
we will consider in the following only situations where all inertial terms
are neglected. However, $\gamma$ is in this case nonzero and arises from the
intrinsic dynamics of the system.

In the absence of external forces, the system is stable for $r > 0$. 
It shows damped oscillations or is overdamped, 
and relaxes to $x=0$.  For $r=0$,
the system becomes unstable and undergoes a Hopf-bifurcation.  For
$r<0$ spontaneous oscillations are generated.  In this regime
the nonlinearity characterized by $\Lambda$ is essential to stabilize
the system and to determine the oscillation amplitude.  In the
periodic limit cycle for $f_{\rm ext}=0$, we can write
\begin{equation}
    x(t)=\sum_{n}x_{n}e^{i n\omega t}
\end{equation}
as a Fourier sum with oscillation frequency $\omega$. Close to the
bifurcation point, i.e. for small but negative $r$, the first 
Fourier mode
$x_{1}$ dominates and obeys to lowest order
\begin{equation}
    {\cal A} x_{1} + {\cal B} \vert x_{1}\vert^2 
    x_{1} =0 \label{eq:bif}
\end{equation}
Here, higher modes $x_{n}\sim x_{1}^n$ are neglected and the complex coefficients 
are given by ${\cal A}=k-\gamma \omega^2
+i\omega r$ and  ${\cal B}=3i \Lambda \omega^3$. Spontaneous 
oscillations occur with frequency $\omega=\omega_{c}\equiv(k/\gamma)^{1/2}$.
This is the only choice for which Eq. (\ref{eq:bif}) has a solution
for which the oscillation amplitude
\begin{equation}
    \vert x_{1}\vert^2 =-\frac{\cal A}{\cal 
    B}=-\frac{r\gamma}{3\Lambda k} \label{eq:amplit}
\end{equation}
is real and positive. The bifurcation point $r=0$ 
is characterized by the
condition that ${\cal A}=0$ vanishes at the critical frequency
$\omega=\omega_{c}$.
If this system is subject to an external force $f_{\rm ext}=2 f_{1} 
\cos(\omega_{\rm ext} t)$ with frequency $\omega_{\rm ext}$, there
are two frequencies in the system, the spontaneous frequency and the
externally imposed one. For simplicity, we focus on the situation where
only one frequency is present, i.e. the system is either outside the
spontaneously oscillating regime ($r \geq 0$), 
or both frequencies are the same
$\omega=\omega_{\rm ext}$.
In this case, Eq. (\ref{eq:bif}) becomes simply
\begin{equation}
    {\cal A} x_{1} + {\cal B} \vert x_{1}\vert^2 
    x_{1} = f_{1} \label{eq:fbif}
\end{equation}

\begin{figure}    
 \centerline{\psfig{figure=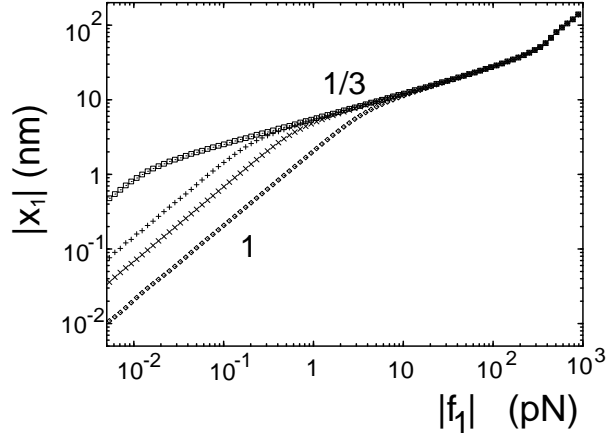,width=8cm}}
 \caption{Fourier amplitude $\vert x_{1}\vert $ of the response of a 
 dynamical system at a Hopf bifurcation to a stimulus force of
 amplitude $\vert f_{1}\vert$ at different frequencies. The data is
 obtained by a numerical solution of the model described in the 
 subsequent section, see \protect\cite{cama00}. The power law of the
 response is indicated.}
\end{figure}

This Equation is generic in the sense that all other terms can be
neglected if the system is sufficiently close to its bifurcation
point.  We therefore focus on the case where $r=0$ and the system is
exactly at the bifurcation.  Two different regimes of the response to
an oscillating force can be distinguished, see Fig.1.  If the applied
frequency is close to the critical frequency, the linear term can be
ignored and
\begin{equation}
    \vert x_{1}\vert \simeq \vert {\cal B}\vert^{-1/3}\vert
    f_{1}\vert^{1/3} \simeq\frac{\vert f_{1}\vert^{1/3}}{ 3^{1/3}
    \Lambda^{1/3}\omega_{c}} \label{eq:nonlin}
\end{equation}
If the frequency mismatch is larger, 
$\vert\omega-\omega_{c}\vert\gg \vert 
f_{1}^{2/3} \vert\;\vert {\cal 
B}\vert^{1/3}\gamma\vert\omega+\omega_{c}\vert$, 
the cubic term is unimportant and the response is linear
\begin{equation}
    \vert x_{1}\vert\simeq  \vert {\cal A}\vert^{-1} \vert 
    f_{1}\vert= \frac{\vert 
f_{1}\vert}{\gamma\vert\omega^2-\omega_{c}^2\vert}\label{eq:lin}
\end{equation}
The response of the system given by Eq. (\ref{eq:fbif}) together with
the oscillation amplitude (\ref{eq:amplit}) characterize the main
properties of an oscillator near the bifurcation point. This approach
can be generalized to situations where more than one frequency are
present \cite{juli01}.

\section{Oscillations generated by molecular motors}

How can a system of the type described in the last section be
realized using biological materials? A spontaneously oscillating 
system must be active, i.e. it has to consume energy from some input. 
Furthermore the system must be able to generate motion and active 
displacements. A prototype system for motion generation in cells are
molecular motors which are enzymes that are driven by a chemical reaction
(usually the hydrolysis of ATP) and which are able to generate 
motion and mechanical work. An individual motor behaves stochastically
and generates on average motion in a preferred direction
along a polar track filament. Because of the stochastic function of
individual molecules, phase coherent oscillations can only be generated
if many motors are involved and their fluctuations become unimportant
for the collective behavior of the entire system.

\begin{figure}
 \centerline{\psfig{figure=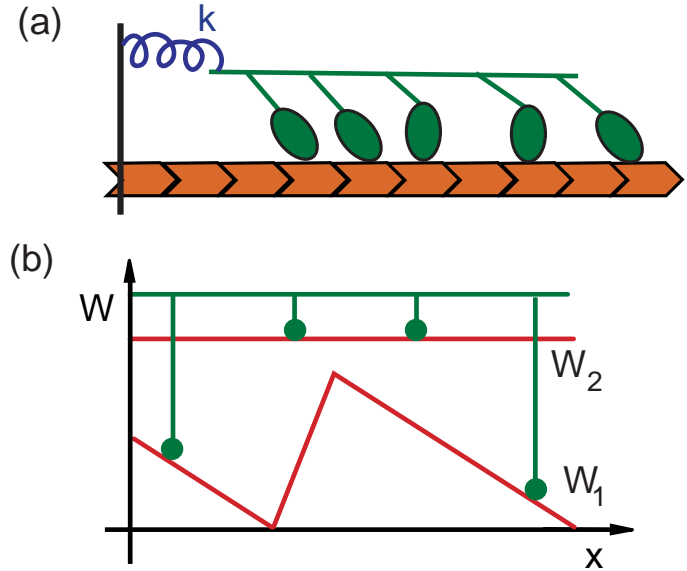,width= 9cm}}
 \caption{(a) Schematic representation of many motors collectively
 working against a spring of elastic modulus $k$. (b) Two state
 model for many motors. Each state is characterized by an energy 
 landscape $W_{i}(x)$.}
\end{figure}

We consider here a specific situation where a large number of molecular 
motors which are rigidly connected and coupled to elastic elements can 
undergo a Hopf-bifurcation \cite{juli97}, see Fig. 2 (a). This description is based
on simple models for the force generation of molecular motors 
\cite{ajda92,magn93,pros94,pesk94,juli97,astu97} generalized to
situations where many motors are coupled \cite{huxl57,juli95}. In the
limit of a very large number of motors $N$, fluctuations in the motor
function become irrelevant and the system can be described by simple
mean-field equations \cite{juli99}. We consider motors
moving along a periodic linear structure of period $\ell$, see Fig. 2 (b). 
In the context of a two-state model for the motors, we study the 
probability density $P(\xi,t)$ to find a motor
at position $\xi$ in state $i=1,2$, which is normalized, 
$P_{1}+P_{2}=\ell^{-1}$, and satisfies
\begin{eqnarray}
      \partial_{t}P_{1}&=&-v\partial_{\xi}P_{1}-\omega_{1}P_{1}+\omega_{2}P_{2}
      \nonumber \\
      \partial_{t}P_{2}&=&-v\partial_{\xi}P_{2}+\omega_{1}P_{1}-\omega_{2}P_{2}
\end{eqnarray}
Here, $\omega_{1}$ and $\omega_{2}$ are the rates of chemical 
transition between the two conformations of the motors.
The velocity $v=\lambda^{-1} (f_{\rm ext} + f_{\rm mot}-k x)$ is generated
by the sum of the externally applied force per motor $f_{\rm ext}$, 
the average force 
\begin{equation}
    f_{\rm mot}=-\int_{0}^\ell d\xi (P_{1}\partial_{\xi 
}W_{1}+P_{2}\partial_{\xi}W_{2})  
\end{equation}
generated per motor and an elastic force in presence of an elastic
element of modulus $k$ per motor. The total friction coefficient per 
motor is denoted $\lambda$. The energy landscapes $W_{i}$ 
characterize the interaction with the filament and are
usually considered to be periodic with period $\ell$.  Using the
condition $\dot x=v$, the system
can undergo a Hopf-bifurcation in the vicinity of which
Eq. (\ref{eq:fbif}) is satisfied with linear and nonlinear coefficients
\begin{eqnarray}
    {\cal A}(\omega,C) &=&i\omega \lambda+k-i\omega\int_{0}^\ell d\xi 
    \partial_{\xi} R \frac{\partial_{\xi}(W_{1}-W_{2})
    }{\alpha+i\omega}\nonumber\\
    {\cal B}(\omega,C) &=&i\omega^3(F_{1,1,-1}+F_{1,-1,1}+F_{-1,1,1})\label{eq:A}
\end{eqnarray}
where
\begin{equation}
    F_{klm}=\int_{0}^\ell d\xi 
    \left(\partial_{\xi}\frac{1}{\alpha+ik\omega}\left(\partial_{\xi}
    \frac{\partial_{\xi} R}{\alpha+il\omega}\right)
    \right)\frac{\partial_{\xi}(W_{1}-W_{2})}{\alpha+im\omega}
\end{equation}
Here, $\alpha(\xi)\equiv\omega_{1}(\xi)+\omega_{2}(\xi)$, 
$R\equiv\omega_{2}/\ell\alpha$, and we have assumed for simplicity that
the potentials and transitions are symmetric with respect to 
$\xi\rightarrow -\xi$. The coefficients ${\cal A}$ and ${\cal B}$ are 
functions of frequency and also depend on other model parameters such
at the transition rates $\omega_{1}$ and $\omega_{2}$. We introduce
a control parameter $C$ which summarizes changes in the transition 
rates which keep $\alpha$ constant and which will be varied in order
to cross the bifurcation point. $C$ could represent the 
concentration of any agent that influences chemical rates, in 
the case of motor molecules for example those of ATP or Ca$^{++}$. 

In order to discuss these somewhat uninstructive expressions, we
consider the simpler case where $\alpha$ is a constant.  In this case,
we find
\begin{eqnarray}
{\cal A} &=&k-\gamma(\omega) \omega^2+i\omega r(\omega)
\nonumber\\
{\cal B} &=&-3iK^{(\rm nl)}_{\rm eff}\frac{\omega^3 
}{(\alpha^2+\omega^2)(\alpha+i\omega)}\label{eq:A0}
\end{eqnarray}
Eq. (\ref{eq:A0}) corresponds to Eq. (\ref{eq:vdp}), however 
with coefficients that depend on both the frequency and the control
parameter $C$
\begin{eqnarray}
    \gamma(\omega) &=&\frac{K^{(2)}_{\rm 
    eff}}{\alpha^2+\omega^2}\nonumber\\
    r(\omega) &=&\lambda-\frac{K_{\rm eff}^{(2)}\alpha}
    {\alpha^2+\omega^2}
\end{eqnarray}
The effective Hookian and nonlinear elasticities of the cross-bridges
\begin{eqnarray}
    K^{(2)}_{\rm eff}(C)&=&\int_{0}^\ell d\xi\; 
R\;\partial_{\xi}^2(W_{1}-W_{2})\\
    K^{(4)}_{\rm eff}(C)&=&\int_{0}^\ell d\xi\; 
R\;\partial_{\xi}^4(W_{1}-W_{2})
\end{eqnarray}
which are functions of $C$ and characterize the second and
fourth derivative of the effective potential of a motor in the bound
state.  Note that the elasticity $K^{(2)}_{\rm eff}$ together with the
chemical transitions generates an effective "mass" $\gamma(\omega)$
and a negative contribution to friction.

If the control parameter $C$ is varied, the system undergoes for a 
critical value $C_{c}$ a
Hopf-bifurcation. This happens as soon as $K^{(2)}_{\rm eff}$ takes a value
for which ${\cal A}(\omega_{c},C_{c})=0$ vanishes for a critical value
of the frequency $\omega_{c}$. In order to characterize this point, 
note first that the imaginary part of $\cal A$ vanishes if 
$r=0$, or $\lambda=\alpha\gamma $. 
The critical frequency at the bifurcation 
satisfies $\omega_{c}=(k/\gamma(\omega_{c}))^{1/2}$ 
and is thus given by  
\begin{equation}
    \omega_{c}=\left( \frac{\alpha k}{\lambda}\right)^{1/2} \quad .
\end{equation}
The range of frequencies that can be obtained by this mechanism
depends on the range of effective elasticities $K^{(2)}_{\rm eff}$
which can result from motor-filament interactions. 
If the motor is detached most of the time
this elasticity is small $K^{(2)}_{\rm eff}\simeq 0$, while in the
other extreme of a motor that rests most of the time in rigor will
generate a $K^{(2)}_{\rm eff}\simeq K_{\rm cb}$ where $K_{{\rm cb}}$
is the cross-bridge elasticity. The highest frequencies the system
can attain at the bifurcation are thus determined by $K_{\rm cb}$:
\begin{equation}
    \omega_{\rm max} \simeq \left(\frac{\alpha K_{\rm cb}}{\lambda}
    -\alpha^2\right )^{1/2}\quad .
\end{equation}

In the limit of a large number of motors discussed here, the generated
oscillations are phase coherent and fluctuations have been
neglected.  If the number of active elements is finite, the
stochasticity of the chemical transitions leads to a stochastic
component in the motion.  In addition to this noise resulting from
active processes, thermal fluctuations can also become relevant. 
These resulting noisy oscillations have a finite time of phase
coherence \cite{juli97,cama00}. 
While a small number of motors could already generate
noisy oscillations with short coherence time, a real Hopf-bifurcation
with phase coherent oscillations is only possible as a collective
phenomenon of a large number of active elements as described here.

\section{Self-organization of motors and filaments}

Cilia and flagella which contain an axoneme are important examples of
mechanically oscillating biological structures.  In these hair-like
appendages of many cells, microtubule doublets are arranged in a 
cylindrical fashion and undergo bending deformations as a result of 
relative forces exerted by a large number of dynein molecular motors 
which are densely attached to the surface of microtubules and act 
on a neighboring microtubules \cite{lind97}, see Fig. 3. At the basal end, 
microtubules are connected in order to prevent global relative sliding.
Internal stresses generated by the motors are thus directly coupled to
bending deformations of the elastic microtubule doublets which are
all bundled in parallel.

\begin{figure}
 \centerline{\psfig{figure=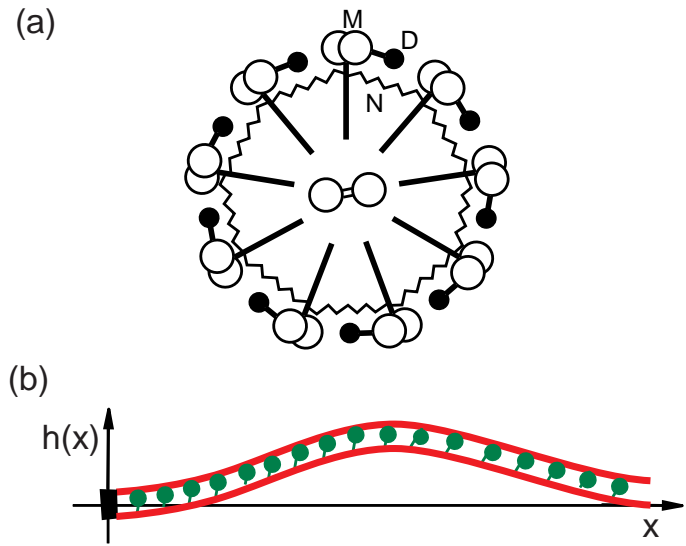,width= 9cm}}
 \caption{(a) Schematic representation of the axoneme.  Nine
 microtubule doublets (M) are arranged in a cylindrical fashion around
 a pair of central microtubules.  Dynein motors (D) are attached to
 the microtubules and exert forces on their neighbors. Elastic 
 elements such as nexins (N) are also present. (b) Two-dimensional
 representation. Two parallel elastic filaments are connected at one 
 end. Active elements exert internal forces. For small amplitudes,
 the shape can be characterized by the deformation $h(x)$ as a 
 function of length.}
\end{figure}

Axonemes are able to generate periodic deformations and to propagate
bending waves along the elastic cilium. These systems fall thus in
the class of systems where many motors coupled to elastic filaments
generate oscillations and spatio-temporal deformation patterns.
Basic physical properties of these structures can 
be captured by considering a
two-dimensional version of this system 
\cite{brok75,hine78,brok85,cama99,cama00b}.  Two elastic filaments are
arranged in parallel at fixed distance $a$.  At one end both filaments
are rigidly connected, everywhere else active elements such as
molecular motors induce locally relative sliding of filaments.
As motors in this situation generate an increasing displacement, the
filament pair bends and generates an elastic force which opposes
filament sliding. This situation therefore corresponds to the case
discussed in the last section where motors work against an elastic 
element. Here, this elasticity is provided by the bending elasticity
of filaments. If this system undergoes a Hopf bifurcation, the whole
filament shape oscillates and exhibits spatio-temporal 
deformation patterns. 

For small deformation amplitudes, the equations of motion of the 
active filament pair can be discussed 
in Monge representation, where the perpendicular displacement $h(s)$
as a function of arclength $s$ with $0<s<L$ is considered. 
To linear order in $h$
we find, \begin{equation}
    \xi_{\perp} \partial_{t} h = -\kappa \partial^4_{s} h -a\partial_{s} f
\end{equation}
where we have assumed for simplicity local viscous friction with 
coefficient $\xi_{\perp}$ and 
where $f(t,s)$ is the internal force per unit length exerted by the motors.
If the coupled motor-filament system undergoes a Hopf-bifurcation,
the selected modes of periodic filament deformations are generic
and can be calculated from very few assumptions as long as the
system is sufficiently close to the bifurcation such that deformation
amplitudes remain small \cite{cama00b}.

In this case, we can use Eq. (\ref{eq:fbif}) to lowest order 
to express the relation $f_{1}(s)\simeq \rho {\cal A} \Delta_{1}(s)$, between the
Fourier amplitude of local sliding displacements
$\Delta=a(\partial_{s}h(s)-\partial_{s}h(0))$ and the amplitude 
$f_{1}(s)$ of the density of shear forces, 
where $\rho$ is the linear density of motors along the 
filaments.  Therefore,
\begin{equation}
    i\omega \xi_{\perp} \tilde h +\kappa \partial_{s}^4 \tilde h
    =-a^2 \rho {\cal A} \partial_{s}^2\tilde h \label{eq:beat}
\end{equation}
together with boundary conditions characterizes the bending patterns
selected at a Hopf-bifurcation with critical frequency $\omega$.
In this equation,
the dynamic linear response function of the motors, ${\cal A}$, plays the role
of a complex eigenvalue. At a Hopf bifurcation, ${\cal A}$ must take
one of a discrete set of complex values. This approach
leads to bending waves which propagate along the filament in a 
direction which depends on the imposed boundary conditions \cite{cama00b}.
These patterns only depend on the solvent viscosity and the bending 
rigidity of microtubules and for given oscillation frequency are
independent of the physical details of the mechanism that generates the
forces.

The frequency selected by the system however is more difficult to
obtain. It depends on properties of the active and passive elements 
which generate internal shear forces. Two different regimes have to be
distinguished: For long filaments complex bending waves are propagated
at rather low frequency. For shorter filaments, there exists a simpler
regime where filaments vibrate without significant wave propagation.
In this latter case, one can ignore to good approximation the detailed shape
of the bending patterns. And  characterize the deformations simply by 
a typical deformation amplitude $h$. We can now think of a situation
where an external force of amplitude $F_{1}$ acts on the tip of the
vibrating cilium and express the linear response
as $h_{1}\sim {\cal A}_{\rm eff}
F_{1}$. Approximating the gradient terms in Eq. 
(\ref{eq:beat}) by their scaling behavior we can write approximately
\begin{equation}
{\cal A}_{\rm eff}\simeq  i\omega \xi_{\perp}L+\kappa/L^3+a^2\rho 
{\cal A}/L
\end{equation}
Using Eq. (\ref{eq:A0}) to express ${\cal A}(\omega,C)$, this leads to
\begin{equation}
    {\cal A}_{\rm eff}=k_{\rm eff}-\gamma_{\rm eff} \omega^2 
    +i\omega r_{\rm eff}
\end{equation}
where
\begin{eqnarray}
    k_{\rm eff} &\simeq&\kappa/L^3 +a^2\rho k/L \nonumber\\
    \gamma_{\rm eff}& \simeq& a^2\rho\gamma(\omega) /L\nonumber\\
    r_{\rm eff}&\simeq &a^2\rho r(\omega)/L+\xi_{\perp}L
\end{eqnarray}
which allows us to estimate the critical beating frequency of the
active filament pair as a function of parameters in several regimes. 
For example in the case where $\kappa_{\rm eff}\simeq \kappa/L^3$ and
$\xi_{\perp}\gg a^2\rho\lambda/L$ we find
\begin{equation}
    \omega_{c}\simeq \left(\frac{\alpha\kappa}{\xi_{\perp}}\right)^{1/2}
    \frac{1}{L^2}\label{eq:wclength}
\end{equation}
This result represents a regime in which the system generates 
oscillations at frequencies which increase significantly for decreasing
length if all other parameters are constant. Furthermore, in this
simple regime, the frequency is completely determined by the
bending rigidity of microtubules $\kappa$, the viscosity of the 
solvent $\xi_{\perp}$ and a typical ATP cycling rate $\alpha$ of the motors.

\begin{figure}
 \centerline{\psfig{figure=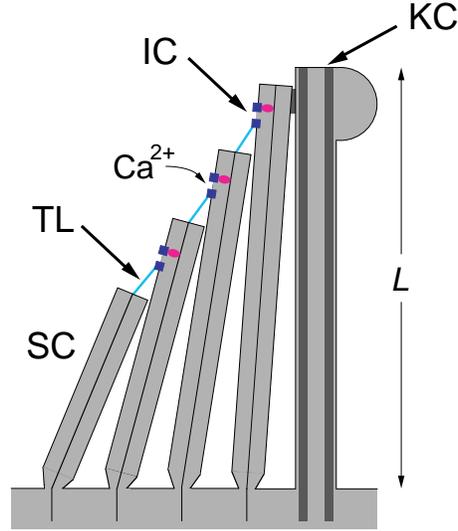,width=6cm}}
 \caption{Schematic representation of the hair bundle of an auditory
 hair cell. A bundle of stereocilia (SC) containing actin filaments 
 form the hair bundle. They are coupled at their tips by fine filaments 
 called tip-links (TL). The kinocilium (KC) is present in the hair 
 bundles on non-mammalian vertebrates and contains an axoneme. 
 Mechano-sensitive ion channels (IC) open upon a bundle deflection 
 and let to an influx of K$^+$ and Ca$^{++}$.}
\end{figure}

\section{Oscillations in sensory systems: hearing}

The cochlea of the inner ear contains about 16000 specialized sensory
cells, called hair cells, which are able to detect sounds at a range
of frequencies from $50Hz$ to $40000$Hz.  The auditory system has an
extraordinary dynamic range which covers $120$dB or 12 orders of
magnitude in sound intensity.  Each hair cell has a characteristic
frequency at which it is most sensitive.  Hair cells are characterized
by a bundle of about 50 finger-like structures called stereocilia
which have a length of 1-10$\mu$m and a diameter of 300nm.  The
stereocilia consist predominantly of bundles of actin filaments that
are surrounded by a membrane which contains mechano-sensitive ion
channels.  Upon a tiny shear deformation of the hair bundle of less
than a nm, these channels open and the subsequent influx of Potassium
and Calcium induced changes in the membrane potential \cite{huds94} (Fig. 4).

The mechanical response of the basilar membrane, which is the
structure inside the cochlea that contains the hair cells, has been
measured as a function of frequency and amplitude of sound stimuli
\cite{rugg92}.  These and other observations have let to a revival of
an idea of Thomas Gold, proposed about 50 years ago \cite{gold48},
that the basilar membrane of the inner ear is not a passively
resonating filter as assumed by the classical theory of hearing
developed by Van Bekesy and others \cite{vanb60}.  Instead, Gold
suggested that active, energy consuming systems are required by the
ear in order to explain some of its extraordinary properties such as a
sharp frequency tuning.
The mechanical response of the basilar membrane reveals a distinct
nonlinear behavior which has all the characteristic properties of a
dynamical system placed exactly at a Hopf bifurcation as described in
section 2 \cite{cama00,egui00}.  Most remarkable is the nonlinear
response to stimuli at the critical frequency $\omega_{c}$, see Eq. 
(\ref{eq:nonlin}) which implies a diverging gain $\vert
x_{1}\vert/\vert f_{1}\vert\sim \vert f_{1}\vert^{-3/2}$ for small
stimulus amplitudes.  This nonlinear behavior of the gain and a
dramatic increase at small amplitudes has been observed in hearing
\cite{rugg92}.  This actively enhanced gain together with a power-law
nonlinearity provides extraordinary sensitivity and sound detection
over a large dynamic range of 6 orders of magnitude in amplitude and
12 orders of magnitude in intensity.

While the properties of a Hopf bifurcation can explain the main
properties of sound detection by the ear, it raises a number of
important questions.  In particular, in order to exhibit a nonlinear
response for small amplitudes, the system has to be tuned with high
precision to its critical point.  This requires a fine-tuning of
parameters which raises doubts as to whether a living cell can profit
from the special properties at a critical point in a reliable way.

A simple and general mechanism to maintain a dynamical system at a
point of operation close to the bifurcation point can be achieved by a
feedback regulation of the control parameter \cite{cama00,cama01}.  This
self-tuning implies that the control parameter is regulated towards
the instability as long as the system is not spontaneously
oscillating, while it is automatically stabilized as soon as
oscillations are detected.  This self-tuning works best in the absence
of external stimuli, when highest sensitivity is needed, 
by adapting the control parameter $C$ as a
function of the detected amplitude of hair-bundle deflections.

\begin{figure}
 \centerline{\psfig{figure=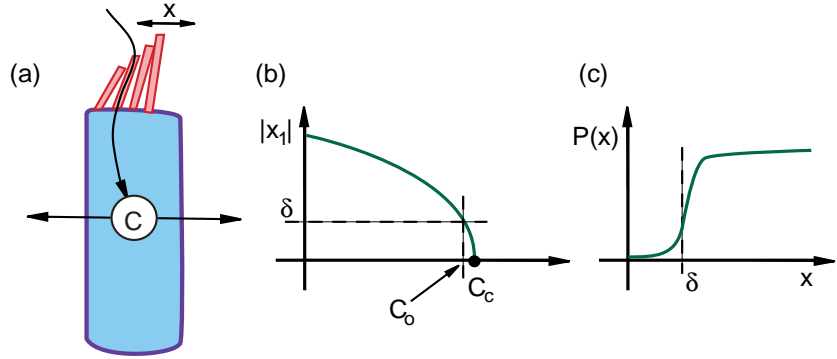,width=11cm}}
 \caption{Simple self-tuning mechanism (schematic).  (a) Regulation of
 the control parameter $C$ associated with the concentration of ions
 such as Ca$^{++}$ which enter the hair cell vie transduction
 channels.  A permanent outflux drives the system towards the
 oscillating side of the bifurcation, influx of ion via transduction
 channels provides a stabilizing feedback.  (b) Fourier amplitude
 $\vert x_{1}\vert$ of spontaneous oscillations as a function of the
 control parameter near the bifurcation point $C_{c}$.  Self-tuning
 brings the system to an operating point $C_{o}$.  (c) Opening
 probability $P(x)$ of ion channels as a function of the deflection
 amplitude $x$.  A signal is generated for deflections larger than
 $\delta$.}
\end{figure}

An illustrative example for self-tuning is achieved in a situation where
the Ca$^{++}$ concentration in the hair cell plays
the role of the control parameter at the bifurcation (Fig. 5). Since
hair-bundle deflections lead to an influx of Ca$^{++}$ into the hair 
cell, this provides for a regulation of the control parameter $C$ of the form
\begin{equation}
    \frac{dC}{dt}=-\frac{C}{\tau}+J P(x)
    \label{eq:selftuning}
\end{equation}
where $\tau$ is a relaxation time of the control parameter 
in the absence of hair bundle deflections $x$. This relaxation drives the
system in the oscillatory regime. As soon as deflections $x$ occur,
ion channels open with probability
\begin{equation}
	P(x)=\frac{1}{1+e^{-(x-\delta)/\ell}}
\end{equation}
and each open channel gives rise to an influx $J$ of Ca$^{++}$,
which drives the system towards the non-oscillating regime.
Here we assume $\tau\gg\omega^{-1}$, i.e. changes in $C$ occur on
time-scales long compared to the oscillation frequency.
%Eq.  (\ref{eq:selftuning}) represents a generic self-tuning 
%mechanism which could be realized in
%different specific ways. It can for example be derived from simple
%models for the opening probility of ion channels as a function of 
%deflection.   
The length scale $\delta$, which for hair
bundles is of the order of 0.3-1nm, indicates the smallest deflection
amplitudes at which a signal is generated by the hair bundle and the 
parameter $\ell$ characterizes the sharpness of the response.  

The self-tuning can now be summarized as follows.
In the absence of spontaneous oscillations (and if no external sound stimulus
is present), the control parameter is decreased within a relaxation
time $\tau$.  As soon as the critical point $C_{c}$ is reached, the
system starts to oscillate and the decrease of $C$ is halted as soon
as the typical oscillation amplitude is of the order of magnitude 
$\delta$, $\vert x_{1}\vert\sim\delta$.  Remember that the onset of spontaneous
oscillations in the absence of a force given by Eq. (\ref{eq:amplit})
can be expressed as
\begin{equation}
    \vert x_{1}\vert \simeq 
    \Delta \left (\frac{C-C_{c}}{C_{c}}\right )^{1/2}
\end{equation}
where $\Delta$ is a characteristic saturation amplitude. We can 
therefore estimate how close to the bifurcation point the system
will be tuned via this mechanisms. introducing the distance
$\Delta C=C_{c}-C$ from the bifurcation, the self tuning brings the
system slightly to the oscillating side of the bifurcation with
\begin{equation}
    \Delta C/C_{c}\sim (\delta/\Delta)^2 \quad .
\end{equation}
For a typical hair cell we estimate $\delta\sim 1$nm and $\Delta\sim
100$nm, thus the system can self-tune with 
$\Delta C/C_{c}\simeq 10^{-4}$. 

These estimates for the useful amplitude range of hair bundle
deflections can also explain the dynamic range of hearing.  In the
regime of nonlinear response, the hair cell can map changes in force
amplitude $\vert f_{1}\vert$ that vary by a factor of
$(\Delta/\delta)^3\simeq 10^{6}$ onto hair bundle deflections $\vert
x_{1}\vert$ which vary over the usable range of $\Delta/\delta$.  This
range of detectable force amplitudes corresponds to a dynamic range of
120dB.

The concept of self-tuning to a Hopf bifurcation can explain seemingly
disconnected but well-known properties of hearing. In addition to 
providing an explanation for the high sensitivity at one frequency and
a large dynamic range, it can naturally account for what is called 
adaptation and fatigue. Fatigue implies that the sensitivity to weak
stimuli is reduced after a subject is exposed to a loud stimulus which 
is a natural consequence of self-tuning. In the presence of a stimulus,
Eq. (\ref{eq:selftuning}) tunes the system away from the bifurcation 
point where sensitivity is reduced. The recovery of high sensitivity
after a strong stimulus only happens after a relaxation time $\tau$ of the
self-tuning feedback to its operation point.

This theory of hearing by a generic mechanism could apply to
many different animals such as mammals, birds, reptiles and amphibians.
These classes of animals however have different cochleas and different 
types of hair cells and they might thus use different physical systems
to realize a Hopf bifurcation and the self-tuning. In the case of
mammals, there is some evidence that so-called outer-hair cells are
able to generate active motion by contracting the whole cell body
\cite{dall92}. Recently, a protein which could play an active role in
these contractions has been identified \cite{zhen00}. Non-mammalian
vertebrates do not possess outer hair-cells. Active oscillators are
therefore expected to exist within the hair bundle itself. Spontaneous
hair-bundle oscillations of amphibian hair-cells have been observed
and studied in detail \cite{mart99,mart00}.

While the physical mechanism at the origin of hair-bundle oscillations
remains mysterious, the preceding sections show that molecular motors
operating in groups could be responsible for oscillations even at 
frequencies up to 10kHz. It is well established that myosins occur
within the stereocilia and could thus be involved in active movements.
Finally, the hair bundles of non-mammalian vertebrates contain in 
addition to many stereocilia a single cilium which contains an axoneme.
Such a structure is ideally suited to play an active role as an 
oscillator and because of its well-conserved structure, its oscillation 
frequency could be controlled by just varying its length. If a cilium
operates in the regime described by Eq. (\ref{eq:wclength}), using
a typical $\alpha\simeq 10^3$s$^{-1}$, $\xi_\perp=10^{-3}$kg/ms being the 
viscosity of water and choosing $\kappa=4 \times 10^{-22}$Nm$^2$ which is
the bending rigidity of 20 microtubules, the critical frequency is
$\omega_{c}\simeq 2 \times 10^4$s$^{-1}$ for $L=1\mu$m. For $L=10\mu$m,
the frequency drops to $\omega_{c}\simeq 10^2$s$^{-1}$. The physical
mechanisms for oscillations generated by molecular motors in a cilium 
therefore could cover the audible frequency range by using a simple
morphological gradient in the cochlea.

\section{Discussion}

An variety of cells contain structures which are able to spontaneously
generate mechanical oscillations.  The principal examples discussed
here are muscle myofibrils, axonemal cilia and flagella as well as
auditory hair cells of the inner ear.  While the details of how these
oscillations are generated still remain largely unknown, it is
expected that motor proteins which undergo a chemical cycle and
stochastically generate displacements and forces are the active
elements involved.  Molecular motors of the cytoskeleton which
interact with elastic filaments are omnipresent in eucariotic cells
and are likely to be involved in such activities.  While in the case
of myofibrils and axonemes the crucial role of such motors is
established, in the case of hair cells the nature of active proteins
is still quite unknown.

The general principles of how oscillations can be generated using such
active proteins can be studied using simplified descriptions.  Such
studies reveal that spontaneous oscillations occur naturally in
systems where a large number of active elements and elastic elements
interact and form an effective material with active properties.  Of
particular interest is the Hopf bifurcation where the system becomes
unstable with respect to oscillatory behavior.  In the vicinity of a
bifurcation, the system has generic properties that can be understood
and characterized without detailed knowledge of the underlying
mechanisms.  Furthermore, the bifurcation point has extraordinary
response properties which are ideal to be used for a frequency
sensitive detector.  A lot of evidence points indeed to the fact that
the ear has adopted this principal mechanism, aided by a self-tuning
feedback that tunes the oscillators to their critical points.

Phase coherent oscillations and a true Hopf-bifurcation only exist in
the thermodynamic limit where an infinite number of active
force-generators are present.  In a realistic situation with a finite
number of molecules present, oscillations are noisy and phase
coherence is lost on long times.  Oscillations of very few, maybe only
a single dynein motors have been reported \cite{shin98}, which however
are very noisy.  Phase coherent oscillations always result from
collective effects in systems with a large number of degrees of
freedom.  Noise plays an important role for cellular oscillations, in
particular in the case of hearing for the detection of weak sounds
\cite{cama00}.  A hair cell tuned to its operation point in the
absence of an external stimulus will exhibit a spectrum of
fluctuations centered around its critical frequency.  Because of the
active nature of the oscillator, these fluctuations arise partly due
to stochastic activity of active elements in addition to thermal
noise.  As a result of these fluctuations, the hair cell generates
already in the absence of any sound hair-bundle motions which have a
strong fluctuating component.  Consequently, unstimulated hair-cells
already generate action potentials at low rate.  Because of their
stochastic nature these action potentials can be distinguished from
the excitation due to a sinusoidal stimulus.  As soon as a weak
stimulus is present, its main effect is to phase lock the spontaneous
motion.  This phase locking to a stimulus
can be detected before amplitude changes occur.  By this mechanism the
ear can detect sounds which have an effect on the amplitude that is
smaller than the noise.

Oscillatory behaviors are also observed in very different types of systems.
An interesting example are lymphoblasts which usually do not show
oscillatory behaviors. However, if the microtubule network is 
depolymerized using nocodazole, the cell attains a state where the
actin system forms a contractile ring which generates a constriction
of the cell. Interestingly, this ring slowly oscillates between the
two poles of the cell \cite{born89}. The mechanism leading to these oscillations is
not understood. However, it is likely that these oscillations are a
result of a self-organization of the actin cytoskeleton with the help
of myosin motors and maybe other components.

The self-organization of motors and filaments can lead to complex
phenomena in various situations.  Such phenomena have in particular
been studied in filament systems which are driven by
molecular motors that temporarily form active crosslinks between
filament pairs.  In such active filament systems, complex behaviors
such as contractions
and the formation of asters and spirals have been observed
\cite{takiguchi91,nsml97,sewynsl98,ns96,sn98}.  A
phenomenological description of the dynamics of active filament
bundles can be used to describe tension generation and contractions in
non-organized bundles \cite{kj00}.  This description takes into account two
different filament interactions: (i) interactions between parallel
filaments which point in the same direction and (ii) between filaments
that are anti-parallel.  If both interactions are present at the same
time, the system can generate patterns of contracted regions along the
bundle which propagate.  In periodic systems such as contractile
rings, this can lead to oscillatory behaviors of bundle contractions
\cite{kcj01}.

Mechanical oscillations in cells thus occur in a large variety of
different situations and with frequencies which can vary over large
ranges.  Slow oscillations of contractile rings start from several
minutes per period; the highest frequencies of active oscillations in
hair cells could exceed 100kHz, e.g. during high frequency sound
detection in bats and whales.

I thank S. Camalet, T. Duke, K. Kruse and J. Prost for stimulating 
collaborations and A. Ajdari, M. Bornens, H. Delacroix, R. Everaers, 
A.J. Hudspeth, A. Maggs, P. Martin for helpful discussions.

\end{document}